\documentclass[%
 preprint,
 amsmath,amssymb,
 aps,
]{revtex4-2}

\usepackage{graphicx}
\usepackage{dcolumn}
\usepackage{bm}

\begin{document}


\title{Bend losses in flexible polyurethane \\ antiresonant terahertz waveguides}

\author{Alessio Stefani}
\affiliation{
 Institute of Photonics and Optical Science (IPOS), School of Physics, University of Sydney 2006, Sydney, New South Wales, Australia
}%
\author{Jonathan Skelton}%
\affiliation{
 Institute of Photonics and Optical Science (IPOS), School of Physics, University of Sydney 2006, Sydney, New South Wales, Australia
}%


\author{Alessandro Tuniz}
 \email{alessandro.tuniz@sydney.edu.au}
\affiliation{
 Institute of Photonics and Optical Science (IPOS), School of Physics, University of Sydney 2006, Sydney, New South Wales, Australia
}%
\affiliation{
The University of Sydney Nano Institute (Sydney Nano), University of Sydney 2006, Sydney, New South Wales, Australia
}%


\begin{abstract}
The quest for practical waveguides operating in the terahertz range faces two major hurdles: large losses and high rigidity. While recent years have been marked by remarkable progress in lowering the impact of material losses using hollow-core guidance, such waveguides are typically not flexible. Here we experimentally and numerically investigate antiresonant dielectric waveguides made of polyurethane, a commonly used dielectric with a low Young's modulus. The hollow-core nature of antiresonant fibers leads to low transmission losses using simple structures, whereas the low Young's modulus of polyurethane makes them extremely flexible. The structures presented enable millimeter-wave manipulation in centimeter-thick waveguides in the same spirit as conventional (visible- and near-IR-) optical fibers, i.e. conveniently and reconfigurably. We investigate two canonical antiresonant geometries formed by one- and six-tubes, experimentally  comparing their transmission, bend losses and mode profiles. The waveguides under investigation have loss below 1\,dB/cm in their sub-THz transmission bands, increasing by 1\,dB/cm for a bend radius of about 10\,cm, which is analogous  to bending standard $125\,\mu{\rm m}$ diameter fiber to a 1.2 mm radius.
\end{abstract}

\maketitle

\section{Introduction}
Terahertz frequencies (0.1-10\,THz), 
located between the microwave and infrared, 
can be harnessed across a wide range of application areas, including 
molecular~\cite{seo2020terahertz}, gas~\cite{mittleman1998gas} and DNA~\cite{fischer2002far} sensing, 
imaging and security~\cite{liu2007terahertz, tuniz2013metamaterial,Atakaramians2017}, pharmaceutical research~\cite{taday2004applications}, short-haul communication (6G)~\cite{saad2019vision}, and other industrial applications~\cite{tao2020non}. 
Owing to the relative infancy of accessible THz sources and detectors~\cite{van1989terahertz}, devices in this range are relatively underdeveloped, with most systems still relying on free space propagation, in contrast to near-infrared and visible frequencies, where optical fibers and photonic circuits are decades ahead~\cite{sackinger2005broadband, bogaerts2020programmable}. To address this limitation, the past few years have shown rapid development of novel terahertz waveguides, 
e.g., polymer-based~\cite{nielsen2009bendable}, sub-wavelength~\cite{roze2011suspended}, porous~\cite{atakaramians2009thz}, and hollow-core tubes/fibers~\cite{setti2013flexible,bao2015dielectric} We refer the reader to some excellent recent reviews on terahertz fibers~\cite{barh2015specialty, humbert2019optical, islam2020terahertz}, which discuss common structures and the physics underlying their guidance mechanisms in great detail.
 
Several factors limit the performance and widespread deployment of fibers and waveguides at THz frequencies. Most materials exhibit relatively large losses per unit wavelength in the terahertz range~\cite{islam2020terahertz}. As a result, many waveguide designs at THz frequencies rely on hollow-core guidance, where fields only weakly overlap with the lossy cladding material. Examples of such structures include metal waveguides~\cite{mcgowan1999propagation}, which straightforwardly harness the conductor's reflection properties; or dielectric tube waveguides~\cite{bao2015dielectric} of varying sophistication~\cite{setti2013flexible}, which then use antiresonances to guide a (low-loss) leaky mode in the hollow core~\cite{gerome2010simplified}, bypassing the lossy dielectric. Although propagation loss is often seen as the main limiting factor, one drawback of such structures is their high rigidity: many metal-~\cite{lu2010bending} and dielectric-~\cite{lu2014terahertz} waveguides, although bendable~\cite{nielsen2009bendable}, are not particularly flexible, inherently limiting their reconfigurability and integration with existing systems. 

A number of numerical and experimental studies have addressed bend losses in hollow core terahertz waveguides. To quote a few experiments, Bao \emph{et al.}~\cite{bao2015dielectric} measured bend- losses in rigid PMMA hollow tubes, down to 10\,cm bend radii. Setti \emph{et al.}~\cite{setti2013flexible} performed similar experiments using multi-tube anti-resonant waveguides, which were then used for remote THz spectroscopy~\cite{lu2014terahertz}. Flexible metal coated dielectric hollow waveguides have also been considered in a number of reports~\cite{doradla2012characterization, navarro2013terahertz, xie2020300}, but typically have cross sections of the order of millimeters to stay flexible, limiting their performance in the sub-THz range. Recent progress in 3D printing technology has further extended the range of potential THz waveguides~\cite{yang20163d} and circuits~\cite{cao2020additive}. However, owing to the rigidity of the materials typically used in additive manufacturing, and due to the cm-length sample dimensions that are commonly achievable with this technique, individual bends are designed \emph{a priori} and are then connectorized to waveguides via separate modules~\cite{cao2020additive}.

Here, we present a comprehensive study of reconfigurable bend losses in hollow-core waveguides composed of polyurethane, a polymer possessing Young’s modulus that is several orders of magnitude lower than the dielectrics typically used for antiresonant waveguides~\cite{stefani2018terahertz,Fleming:17}. Polyurethane waveguides can possess a bend radius $R_b$ of about 10 times the waveguide diameter and as a result, cm-scale waveguides, necessary for low-loss antiresonant guidance but typically prohibitively rigid, are extraordinarily flexible when made of polyurethane. This property was recently exploited for the generation of orbital angular momentum modes~\cite{stefani2018terahertz}, and is here used to showcase their transmission- and guidance- properties in a number of reconfigurable bends in three dimensions, simply via mechanical manipulation. We believe that such waveguides will add to the growing library of potential structures to be used as a convenient platform for monolithically connecting terahertz components at will, down to cm-scale bends. 

\begin{figure}[b!]
\centering\includegraphics[width=\textwidth]{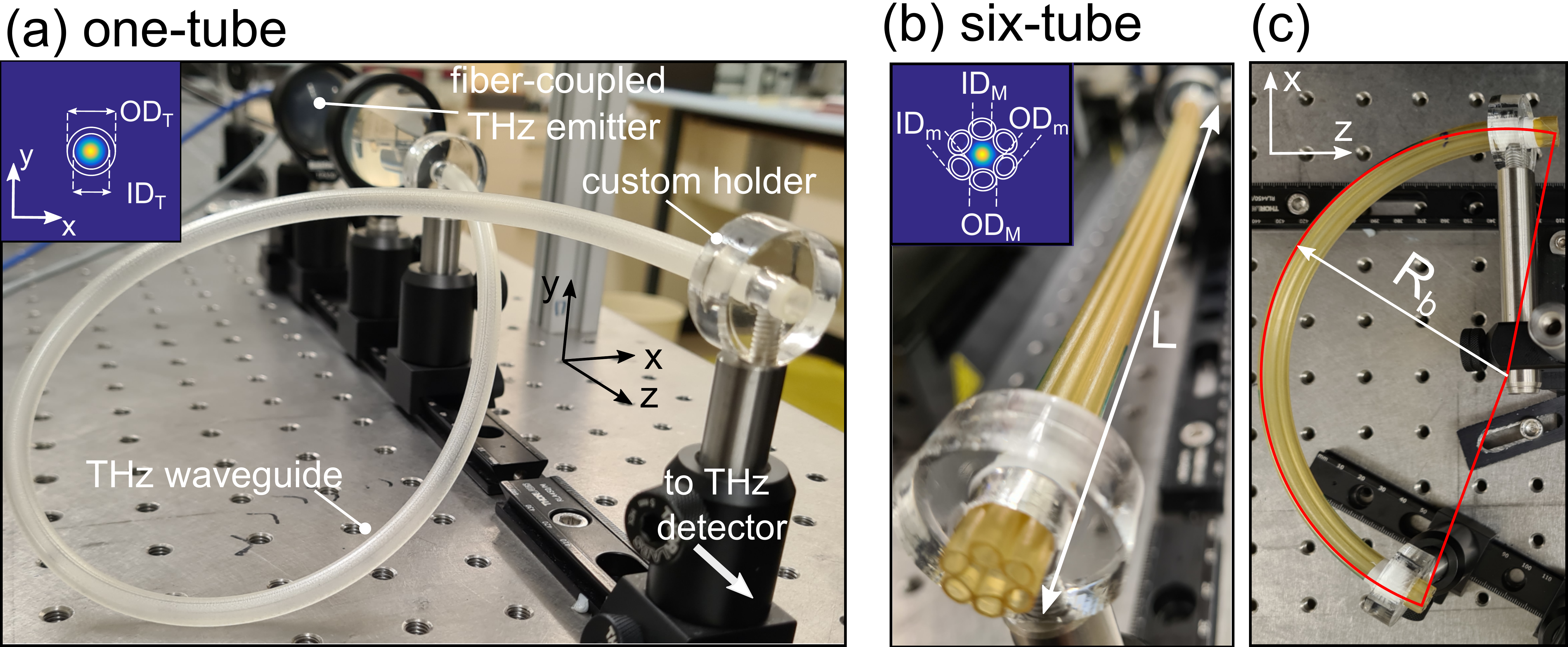}
\caption{Flexible polyurethane antiresonant terahertz waveguide experiment summary and reference frame. We measure and characterize the transmission- and bend- losses of (a) a single-tube waveguide, here termed ``one-tube'' to distinguish it from the (b) ``six-tube'' polyurethane waveguide considered. Inset: calculated modes in the sub-THz mode, highlighting waveguide feature sizes and confinement characteristics (window size: $24 \times 24\,{\rm mm}^2$). Note in particular the high flexibility and reconfigurability of the waveguides. (c) Example bent waveguide configuration, including the bend radius $R_b$. See main text for the numerical values of the geometric parameters labeled.}
\label{fig:fig1}
\end{figure}

\section{Experimental approach}

Figure~\ref{fig:fig1} shows our experimental approach, including a summary of the geometries considered and the experiments performed. We directly compare the propagation and bend characteristics of the two canonical hollow-core antiresonant waveguides: a single dielectric tube, shown in Fig.~\ref{fig:fig1}(a) (here termed one-tube WG); and one of the simplest photonic crystal fibers, shown in Fig.~\ref{fig:fig1}(b), whose cladding is formed by a single layer of hexagonally arranged tubes (here termed six-tube WG).  

Figure~\ref{fig:fig1}(a) shows a photograph of our experimental setup. All antiresonant waveguides are characterized in a commercially available fiber-coupled THz Time Domain Spectroscopy (TDS) system (Menlo TERA K15), with the electric field polarized in $x$ using the reference frame shown in Fig.~\ref{fig:fig1}(a). The fiber-coupled THz system, in combination with custom-built holders that keep the aligned waveguides in place, ensures that the input- and output- conditions are left unchanged while the samples are bent as desired. We perform experiments on both straight waveguides of length $L$ (as shown in Fig.~\ref{fig:fig1}(b) for the six-tube case), and for bent waveguides (as shown in Fig.~\ref{fig:fig1}(c)). In the latter case, the bend radius $R_b$ is trivially obtained from the fiber length and angle swept.

The commercially available polyurethane tube (Grayline LLC) has an outer diameter $\rm OD_{T} = 6.5\,{\rm mm}$ and an inner diameter $\rm ID_{T} = 4.8\,{\rm mm}$, thus possessing a thickness of 0.85\,mm. The six-tube WG is fabricated by stacking six of the single tubes and drawing them under low drawing tension~\cite{PUPatent}. Note that the tubes become elliptical upon drawing: the major- (M) and minor (m) inner diameter (ID) and  and outer diameter (OD) are $\rm OD_M = 3.6\,{\rm mm}$, $\rm OD_m = 3.0\,{\rm mm}$, $\rm ID_M =2.7\,mm $ and $\rm ID_m = 2.1\,{\rm mm}$, leading to a nominal core diameter of 4\,mm and a strut thickness of 0.45\,mm. The two waveguides therefore have a comparable core diameter (to within $< 20\%$), and strut thickness that differs by a factor of two (which result in different antiresonant transmission bands). 
The difference in thickness is an unavoidable consequence of the scaling in the drawing  process. Note that the yellowing of the waveguide, which does not affect the experiments, is due to the thermal history of the fiber, to which the drawing process had only a small contribution.

\section{Experiments}

\subsection{Propagation loss and guided intensity profiles}

\begin{figure}[t!]
\centering\includegraphics[width=\textwidth]{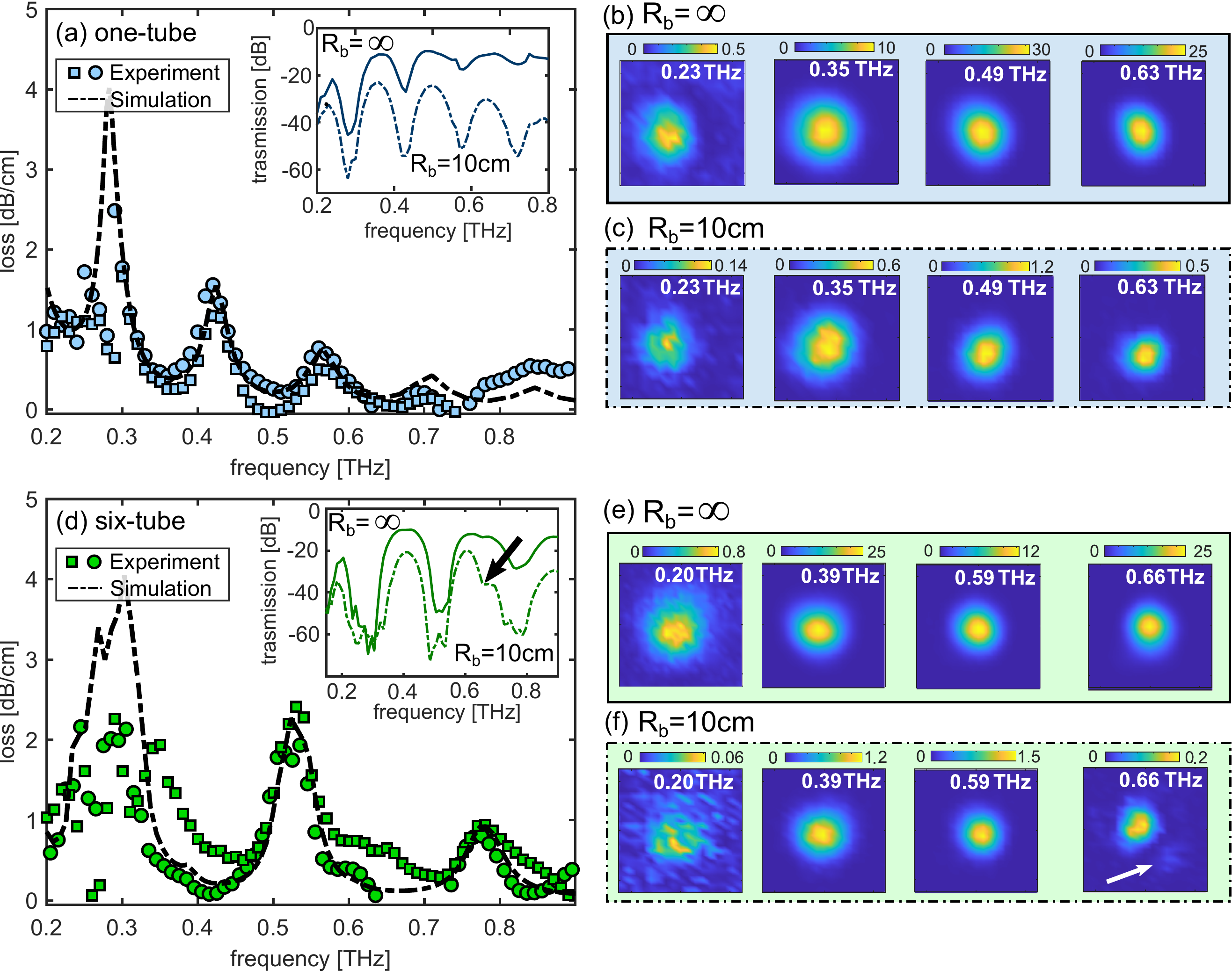}
\caption{Experimentally measured loss- and guiding- characteristics of the waveguides considered. (a) Measured (blue markers) and calculated (dash-dotted line) propagation loss of the one-tube WG as a function of frequency. Inset: example tube transmittance (L=17\,cm) for straight ($R_b = \infty$) and bent ($R_b = 10\,{\rm cm}$) configurations. (b) Experimentally measured intensity distribution in its transmission-band frequencies as labelled, for $R_b = \infty$ and (c) for $R_b = 10\,{\rm cm}$. (d) Measured (green markers) and calculated (dash-dotted line) propagation loss of the six-tube WG.  Inset: example tube transmittance (L=24\,cm) for straight ($R_b = \infty$) and bent ($R_b = 10\,{\rm cm}$) configurations. Black arrow in inset: example resonance feature upon bending (here: 0.66\,THz). (e) Measured intensity distribution at different frequencies as labelled for $R_b = \infty$ and (f) for $R_b = 10\,{\rm cm}$. White arrow highlights leakage into one of the surrounding tubes at resonance (0.66\,THz). All color bars are consistently normalized. Circles and square represent the results obtained from two independent measurements.}
\label{fig:fig2}
\end{figure}

The waveguide attenuation was experimentally obtained by measuring the transmission through three straight waveguides (5--24\,cm lengths) with the same cross-sectional profiles, and fitting it for each frequency with an exponential function. The blue markers in Fig.~\ref{fig:fig2}(a) show the measured loss in the case of a single tube, showing regions of alternating low/high absorption corresponding to the strut's high reflectivity at antiresonance/resonance~\cite{litchinitser2003resonances}. The total coupling efficiency, extrapolated from the measurements at $L=0$, is between 0.5--10\,dB in the transmission bands of this frequency range. A comparison with calculated loss of the fundamental mode, using a finite element modes solver (COMSOL), is plotted as a dash-dotted line, showing excellent agreement. All calculations use a constant refractive index of $n=1.6 + 0.03i$ in the frequency ranges considered, consistently with reported values~\cite{stefani2018terahertz}, and confirmed by preliminary measurements~\cite{jepsen2019phase} using a 1\,mm polyurethane film. Measurement errors at higher frequencies are due to challenges in obtaining a reliable coupling to the fundamental mode, resulting from the highly multimode nature of the WGs in this range.

An example transmittance measurement ($L=17$\,cm) is plotted in the inset of Fig.~\ref{fig:fig2}(a) (solid line), showing the signature regions of high- and low- transmission. The dashed line shows the same measurement when the fiber is bent with $R_b=10\,{\rm cm}$: a frequency-dependent drop in power is observed, with higher frequencies showing more loss, in agreement with previous analyses for rigid polymer tubes~\cite{setti2013flexible, bao2015dielectric}. To complement our transmission measurements, we image the field distribution at waveguide output by repeating the THz-TDS measurements while raster scanning a 1\,mm metal aperture over the waveguide endface~\cite{Adam2011,Kihm:13} (window area: $5 \times 5\,{\rm mm}^2$; lateral step size: $250\,\mu{\rm m}$), collecting the transmitted field at each position of the aperture. Figure~\ref{fig:fig2}(b) and (c) respectively show the mode images for  $R_b=\infty$ and $R_b = 10\,{\rm cm}$, corresponding to the spectrum in the inset of Fig~.\ref{fig:fig2}(a). These measurements highlight that the field remains confined in the hollow core upon bending with only minimal spatial mode deformation. 

The above measurements are repeated for the six-tube WG. Figure~\ref{fig:fig2}(d) (green markers) shows the resulting measured loss. The wider transmission bands are due to the thinner strut thickness, in agreement with simulations (dash-dotted line). Typical transmittance measurements for the straight- and bent- six-tube waveguide, shown in the Fig.~\ref{fig:fig2}(d) inset, already show key differences with respect to the one-tube case. Firstly, despite a comparable bend radius ($R_b = 10\,{\rm cm}$) and the use of a longer waveguide ($L=24$\,cm), the fraction of power lost by the dominant frequency in each transmission band in the six-tube WG is less than for the one-tube WG -- especially at higher frequencies; secondly, we observe an additional resonance feature  within the transmittance windows, highlighted by a black arrow in the Fig.~\ref{fig:fig2}(d) inset. This feature was previously observed, for example, in PMMA six-tube waveguides~\cite{bao2015dielectric}, and  attributed to coupling between the hollow core and the cladding tubes upon bending~\cite{frosz2017analytical}. The corresponding experimental raster-scanned intensity measurements for straight and bent waveguides, shown in Fig.~\ref{fig:fig2}(e) and (f) respectively, confirm that confinement is largely maintained upon bending, and that the resonant feature at $f=0.66\,{\rm THz}$ is associated with coupling of power into the cladding (white arrow in Fig.~\ref{fig:fig2}(f)).

\subsection{Dependence on bend radius}

\begin{figure}[t!]
\centering\includegraphics[width=\textwidth]{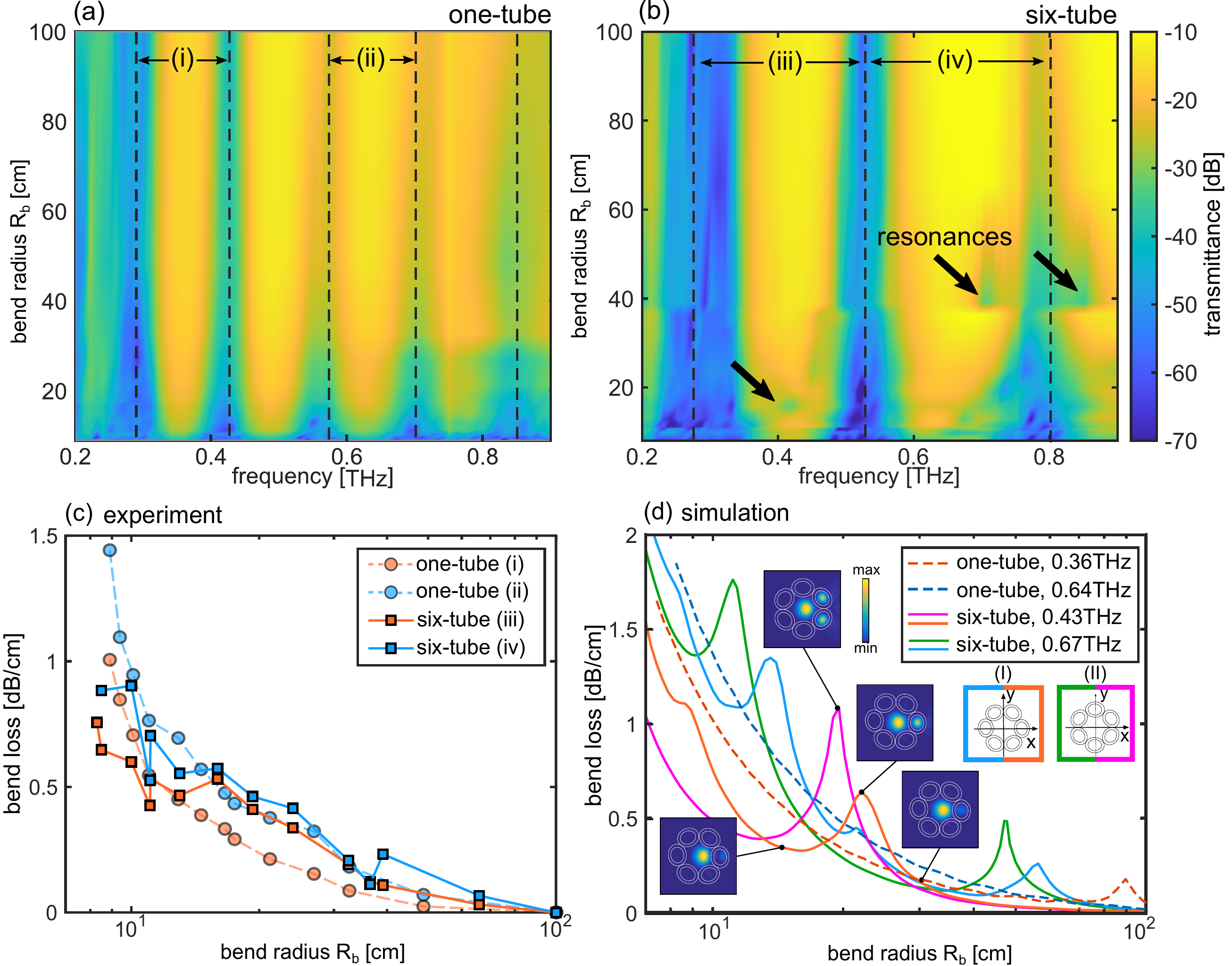}
\caption{Colourplot of the measured transmittance as a function of frequency and bend radius for (a) the one-tube waveguide and (b) the six-tube waveguide of the same length ($L = 24$\,cm). Black arrows highlight example resonant coupling to cladding tube modes. Vertical dashed lines separate the transmission bands: (i) 0.3--0.42\,THz and (ii) 0.57--0.7\,THz for the one-tube WG; (iii) 0.28--0.55\,THz and (iv) 0.56--0.79\,THz for the six-tube WG. (c) Measured loss introduced by bending the waveguide (in dB/cm) obtained by integrating the bands in (i)--(iv) labelled in (a)--(b). Squares: one-tube WG. Circles: six-tube WG. (d) Calculated bend loss for the one-tube (dashed lines) and six-tube (solid lines) WGs using a conformal mapping approach~\cite{heiblum1975analysis}. Blue/orange and green/magenta curves represent six-tube waveguides at two frequencies within the bands in (c), rotated by $30^\circ$ as shown in the inset. Example mode images around resonance show coupling into cladding tube modes upon bending.}
\label{fig:fig3}
\end{figure}

Following these preliminary measurements, we now investigate the bend loss and guiding properties of each waveguide in detail. We start by gradually reducing the bend radius of two waveguides of the same length ($L=24$\,cm), and measuring the transmitted spectrum in each configuration. The transmittance colourplot (in dB) as a function of frequency and bend radius $R_b$ for the one- and six-tube waveguide is shown in Fig.~\ref{fig:fig3}(a) and Fig.~\ref{fig:fig3}(b), respectively. These experiments reveal the sporadic emergence of resonances in all six-tube WG transmission bands, but only at specific values of $R_b$, (black arrows in Fig.~\ref{fig:fig3}(b)), which are absent in the one-tube case. To quantify how the bends impact the transmission band per unit WG length, we integrate the intensity at each bend radius in the two highest-throughput transmission bands for each waveguide as labelled in Fig.~\ref{fig:fig3}(a),(b), and divide by the WG length. The resulting bend loss, in dB/cm, is shown in Fig.~\ref{fig:fig3}(c). We find that, for the one-tube WG (circles in Fig.~\ref{fig:fig3}(c)), higher frequencies exhibiting larger bend losses, which monotonically increase as the bend radius is reduced. In contrast, the six-tube WG (squares in Fig.~\ref{fig:fig3}(c)) shows a loss that fluctuates with decreasing $R_b$ -- being either comparable to or greater than for the one-tube WG for $R_b > 10\,{\rm cm}$. This difference is an immediate consequence of the resonances highlighted by Fig.~\ref{fig:fig3}(b). However, at smaller bend radii the bend loss for the six-tube WG is significantly lower than for the one-tube WG: for the 24\,cm length WGs considered and the lowest-frequency band in each WG, the bend loss in the six-tube WG is up to 10\,dB lower than for the single tube when $R_b = 7\,{\rm cm}$.

To interpret these results, we compare our experiments with numerical models. We utilize a full finite element mode solver (COMSOL) to model an equivalent refractive index profile $n_b$ for the bent WGs, given by~\cite{heiblum1975analysis}

\begin{equation}
n_b(x,y) = n(x,y) \exp(-x/R_b),
\label{eq:nb}
\end{equation}
where $x$ is the (horizontal) bending direction as per Fig.~\ref{fig:fig1}(c), and $n(x,y)$ is the cross-sectional refractive index distribution of the straight waveguide considered. For the center frequency of each transmission band, we calculate the complex effective index of the fundamental mode at each value of $R_b$. The calculated modes' bend losses, expressed in dB/cm, are shown in Fig.~\ref{fig:fig3}(d). For the one-tube case, shown here as dashed lines, the experimental features are well reproduced: the loss increases monotonically, with higher frequencies losing more power upon bending, in agreement with previous experiments on rigid tubes~\cite{bao2015dielectric, setti2013flexible}.

The bend losses  for the six-tube case exhibit richer experimental features, and require careful analysis. Although the six-tube waveguide input is oriented as per Fig.~\ref{fig:fig1}(b), some rotation upon bending is inevitable  under normal laboratory conditions due to the high waveguide flexibility. As a result, in our calculations we consider two limiting cases that are rotated $30^{\circ}$ with respect to each other, as shown in the Fig. 3(d) inset: (I) the middle of the tube intersects the $x$-axis (blue/orange lines); (II) the struts join at the $x$-axis (green/magenta lines). Both configurations are encountered during propagation due to small rotations upon bending, and each produce resonant features at similar bend radii that which are associated with coupling of light from the core mode to the tubes in the cladding, as  shown in the mode colourplots in the inset of Fig.~\ref{fig:fig3}(d). We also find that orientation (I) exhibits lower bend losses than (II); intuitively, configuration (II) possesses twice as many equivalent tube waveguides to couple to upon bending in $x$, leading to higher losses.
Finally, the calculations confirm that the bend losses at smaller bend radii ($R_b<10\,{\rm cm}$) are significantly lower than for the single-tube case. Note that the simulation curves in Fig.~\ref{fig:fig3}(d) show sharper resonant features than the experimental curves in Fig.~\ref{fig:fig3}(b) because the former considers the loss at one frequency only, and the latter over the entire integrated transmission band. 

\begin{figure}[t!]
\centering\includegraphics[width=0.8\textwidth]{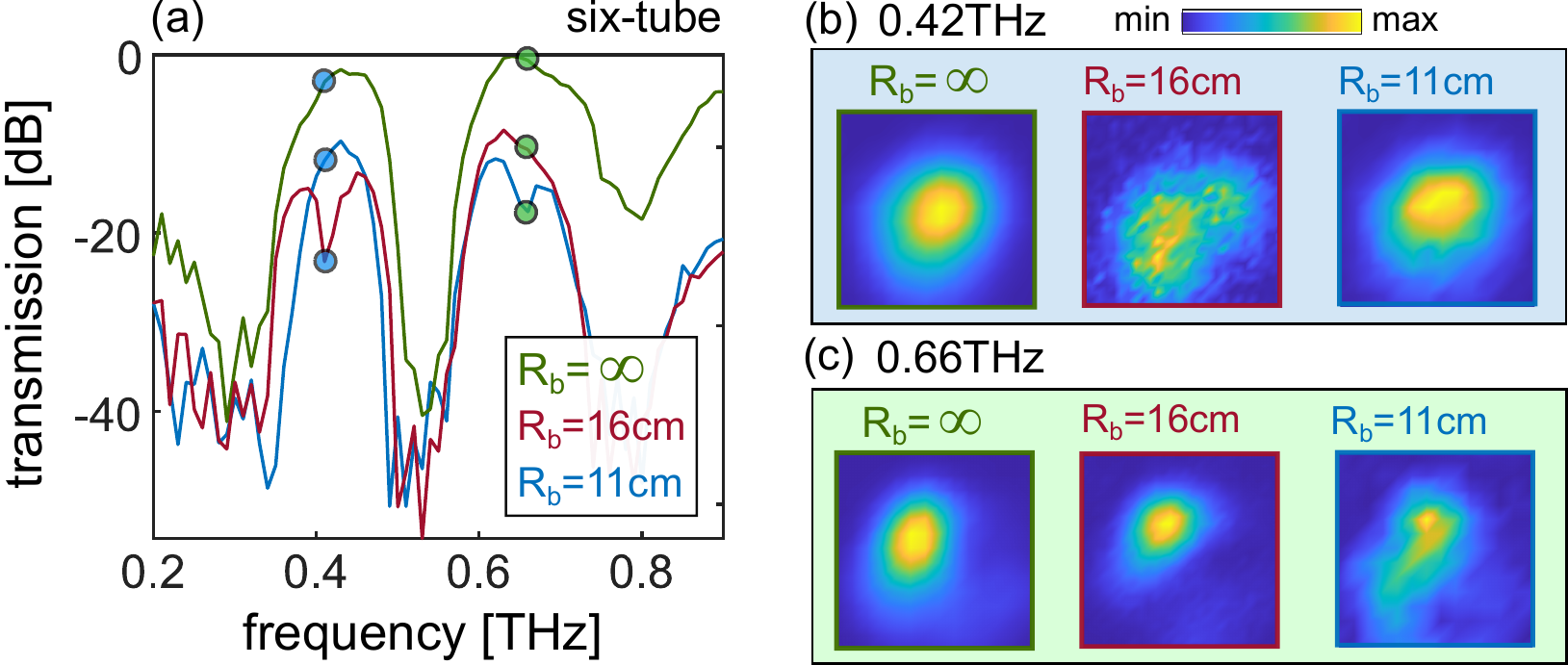}
\caption{Example transmittance measurements for the six-tube waveguide and three bend radii, $R_b = \infty$ (green), $R_b = 16\,{\rm cm}$ (red), $R_b = 11\,{\rm cm}$ (blue), showing resonant features at different bend radii and frequencies. (b) Experimentally measured mode images at 0.43\,THz, showing coupling into a cladding tube mode at $R_b = 16\,{\rm cm}$; at the smaller bend radius of $R_b = 11\,{\rm cm}$, coupling does not occur and the guided intensity profile is comparable to $R_b = \infty$. (c) Experimentally measured mode images at 0.66\,THz, where coupling occurs into a cladding tube mode at $R_b = 11\,{\rm cm}$.}
\label{fig:fig4}
\end{figure}

To elucidate the impact of such resonant features on the WG properties, we measure the guided intensity distribution in two comparable bent WG configurations, whose spectra are shown in Fig.~\ref{fig:fig4}(a). Relative to the straight WG (green curve in Fig.~\ref{fig:fig4}(a)), a bend radius of $R_b = 16\,{\rm cm}$ exhibits a resonant feature at 0.42\,THz which reduces the total transmission significantly below the next transmission band at 0.66\,THz; reducing the bend radius further results in a recovery of the power at 0.42\,THz and the emergence of a resonant feature at  0.66\,THz, in qualitative agreement with the theoretically predicted curves of Fig.~\ref{fig:fig3}(d). The associated experimental images at 0.42\,THz (shown Fig.~\ref{fig:fig4}(b), light blue background) show that the resonance at $R_b = 16\,{\rm cm}$ (middle) is associated with fields coupling into the cladding when compared to the straight waveguide (left), but core guidance is recovered by bending the waveguide further to $R_b = 11\,{\rm cm}$ (right). Similarly, the intensity profiles for the next band centered in 0.66\,THz (shown in Fig.~\ref{fig:fig4}(c), light green background), exhibit the same overall features (i.e., off-resonant core guidance, and the emergence of field in the cladding on resonance.)

\begin{figure}[t!]
\centering\includegraphics[width=\textwidth]{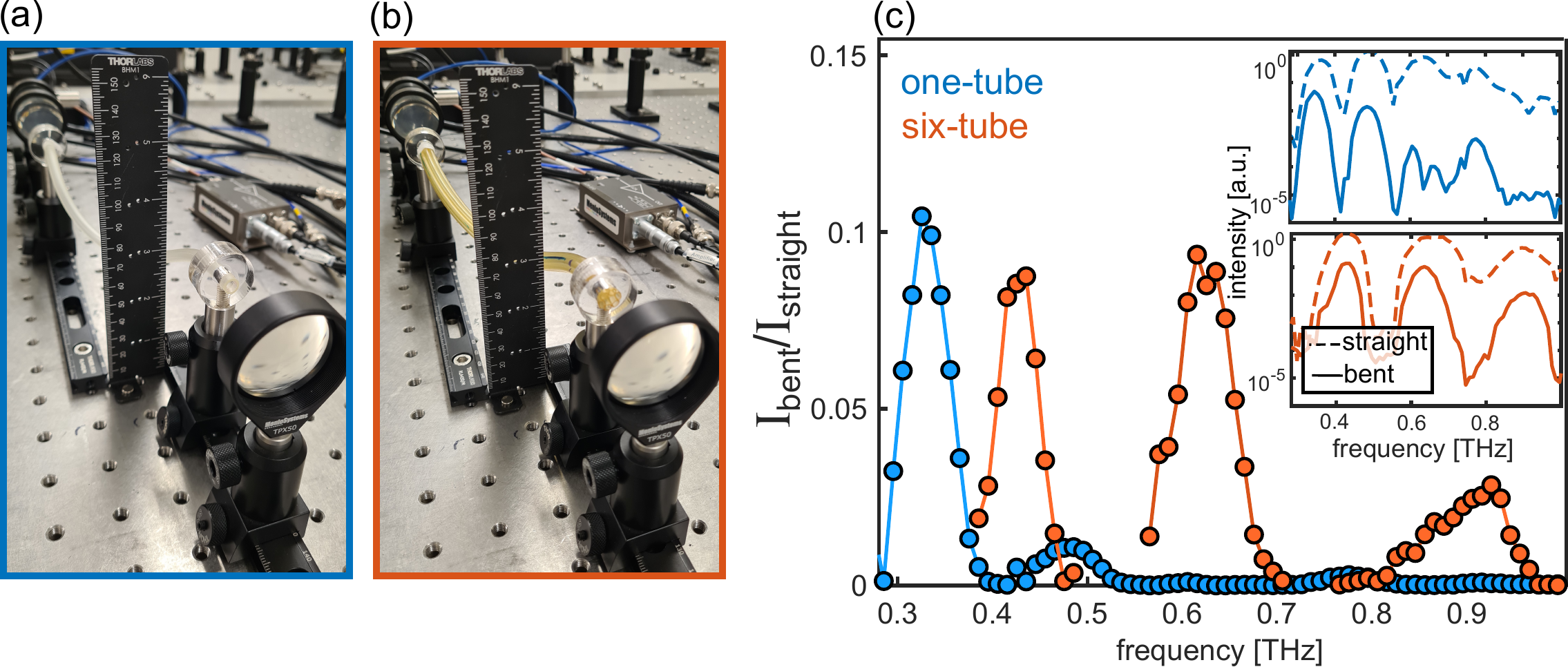}
\caption{Example transmission measurements of two waveguides of the same length, around the same obstacle using a three-dimensional bend, showcasing the reconfigurability and flexibility of our polyurethane antiresonant (a) one-tube and (b) six-tube waveguide. (c) Transmitted intensity by the bent waveguide $I_{\rm bent}$, normalized with respect to that of the straight waveguide $I_{\rm straight}$. For such small bend radii, the six-tube WG transmittance (red) outperforms the one-tube WG (blue) over a broader band. Inset show the associated transmission data for the one-tube and six-tube case, i.e., the transmitted intensity for the straight- (dashed line) and bent- (solid line) waveguides.}
\label{fig:fig5}
\end{figure}

These results clearly suggest that structured antiresonant waveguides are not particularly advantageous over their single-tube counterpart for \emph{larger} bend radii, owing to core-cladding coupling via tube resonances. However, they provide lower bend losses for \emph{smaller} bend radii, an avenue which has been under-explored in this context, due to the high rigidity of commonly used polymers and metals when waveguides of this size are used. Using polyurethane as the device material now enables access to cm-scale arbitrary bends in three dimensions. One example of such a bend is shown in Fig.~\ref{fig:fig5}(a) for a 24\,cm one--tube waveguide, and in Fig.~\ref{fig:fig5}(b) for a six-tube WG of the same length,  corresponding to $R_b < 5\,{\rm cm}$ in a $\sim 15\,{\rm cm}$ WG length. Note that the arbitrary bend in Fig.~\ref{fig:fig5}(a) and \ref{fig:fig5}(b) was chosen to be approximately the same, to the best of our ability, to allow for a fair comparison. The associated transmitted intensity for straight- and bent- WGs (i.e., $I_{\rm straight}$ and $I_{\rm bent}$, respectively) is shown in inset of Fig.~\ref{fig:fig5}(c) as dashed- and solid- lines. Figure~\ref{fig:fig5}(c) also shows the resulting ratio $I_{\rm bent}/I_{\rm straight}$, which presents the most important difference between the one- and six-tube WGs. Although the power lost due to bending in the lower frequency bands is comparable here for both geometries, the six-tube waveguide shows significantly improved transmission for higher frequencies -- up to $\sim 40\,{\rm dB}$ in this particular case.
Structured antiresonant waveguides thus transmit more power over  wider band compared to single tubes for small bend radii, in spite of their slightly bulkier footprint.  These results confirm the potential for polyurethane as a flexible platform for three-dimensional terahertz fiber circuitry, highlighting the benefits of structured tube WGs when extreme bends are used.

\newpage

\section{Conclusions}

In conclusion, we have investigated two canonical hollow-core polyurethane antiresonant waveguides formed by one- and six- tubes. Both WGs are suitable and practical for short-haul sub-THz frequency guidance, showing overall comparable performances with a measured transmission loss below 1 dB/cm in their supported 0.1--1\,THz transmission bands. Such bands/losses are slightly wider/lower for the six-tube WG, due to its thinner struts and stronger antiresonance effects~\cite{pearce2007models}. These WGs further showcase polyurethane as an extremely flexible THz fiber platform, since it allows bend radii as small as ten times the waveguide diameter. Our specific experiments used WGs with $\sim$1\,cm cross-section, enabling guided-wave transmission below 10 cm bend radii for wavelengths up to 3\,mm. To put this into perspective, this is analogous to a standard step-index fiber of $125\,\mu{\rm m}$ diameter guiding a 40\,$\mu{\rm m}$ wavelength within a 1.2\,mm bend radius. Such extreme WG bends introduced a maximum loss increase of $\sim 1$\,dB/cm, with the transmitted field staying confined to the hollow core. While the six-tube waveguide shows resonant coupling to cladding tube modes for specific frequencies when the bend radii are greater than 10\,cm, it outperformed the one-tube waveguide for smaller bend radii.

Finally, we note that the ability to fabricate such waveguides using standard fiber drawing techniques~\cite{tuniz2012fabricating} is critical for producing long waveguide lengths. Further harnessing the extensive amount of ongoing research on antiresonant fibers at higher frequencies is likely to improve their performance and extend their operating frequency, complementing ongoing efforts in fabricating extremely flexible fibers for a wide variety of optoelectronic applications~\cite{leber2019stretchable,shabahangsingle, chen2021elastic,Stefani:20,Runge:18}. In the specific context of terahertz technology, this work demonstrates the potential of antiresonant polyurethane waveguides for making guided waves in this frequency range more practical, flexible, and accessible.

\section*{Funding}
Australian Research Council (DE200101041).

\section*{Acknowledgments}
This work was performed in part at the NSW node of the Australian Nanofabrication Facility. J.S. acknowledges the University of Sydney Faculty of Science Summer Research (Denison) Program.  We gratefully acknowledge fruitful discussions with Simon C. Fleming, Boris T. Kuhlmey, Justin Digweed, and Maryanne C. J. Large. 

\section*{Disclosures}
The authors declare no conflicts of interest.

\bibliography{sample}

\end{document}